# Computing Economic Equilibria by a Homotopy Method

Zoltan Pap

Subotica Tech – College of Applied Sciences, Subotica, Serbia

papzoli@vts.su.ac.rs

*Abstract*— **In this paper the possibility of computing equilibrium in pure exchange and production economies by a homotopy method is investigated. The performance of the algorithm is tested on examples with known equilibria taken from the literature on general equilibrium models and numerical results are presented. In computing equilibria, economy will be specified by excess demand function.**

## I. Introduction

Economic equilibria are usually solutions to fixed point problems [6], and the fixed point theory is the basis of constructive theorems, which give conditions for existence of economic equilibria and provide algorithms for computing these points. These algorithms, called simplicial algorithms, have a finite convergence property. However, equilibria of the large-scale economic problems may be difficult to compute with these algorithms [12]. Alternative methods are developed, which perform well on high dimensional economic equilibrium problems. The fast development of computer science and powerful computers enable equilibria computation of extremely large-scale and nonlinear economic models. Unfortunately, powerful computers are not sufficient for solving these models. Inefficient algorithms can increase the computation time and even provide inaccurate results. As one can see, reliable and good computation algorithms are important ingredients in solving and analyzing economic models. In literature, there are a lot of proposed algorithms which compute equilibria in general equilibrium models. Algorithms based on homotopy method take an important place among this class of computational methods. Homotopy methods for solving fixed point problems were introduced in [2]. As fixed points play an important role in the theory of economic equilibria, homotopy methods have been intensively used in this field. The homotopy method is a class of techniques for solving systems of nonlinear equations. As opposed to Newton's and Newton-like methods, which are locally convergent, homotopy methods are globally convergent starting from an arbitrary initial point.

The market equilibrium problem consists of finding a set of prices and allocations of goods to economic agents such that each agent maximizes her utility, subject to her budget constraints, and the market clears. The equilibrium equations, which are satisfied under mild assumptions [7], express a static condition characterized by the fact that the market demand for each good equals its market supply. These equations are defined by the excess demand function $\xi : \Re_{++}^D \to \Re^D$, where $D$ is finite commodity space, and commodity prices are strictly positive. By allowing that the price of certain good is equal to zero, one obtains a more general definition of the equilibrium. A vector of prices $p^* \in \mathcal{R}_+^D$ is an equilibrium if and only if

$$p^* \geq 0, \quad \xi(p^*) \leq 0, \quad p^{*T}\xi(p^*) = 0. \qquad (1)$$

Problem (1) is known in literature as the nonlinear complementarity problem (NCP).

In this article the possibilities of solving NCP by homotopy methods will be investigated. By deriving the robust homotopy algorithm for solving a NCP, one obtains an efficient computational method for computing economic equilibria. The rest of the paper is organized as follows: Section 2 describes homotopy method; Section 3 is devoted to the basic concept of the equilibrium; Section 4 presents numerical results obtained applying the algorithm on well-known examples, while Section 5 contains the summary and conclusion.

## II. The Homotopy Algorithm

### A. General notes on homotopy methods

Homotopy methods provide a useful approach to find the zeros of smooth mapping $F : \Re^n \to \Re^n$ in a globally convergent way. Such methods have been used to constructively prove the existence of solutions to many economic and engineering problems. The idea is to transform a difficult problem into a simpler one with easily calculated zeros and then gradually deform this simpler problem into the original one computing the zeros of the intervening problems and eventually ending with a zero of the original problem. Deformation of the difficult problem is done by a homotopy function $H : \mathcal{R}^n \times \mathcal{R} \to \mathcal{R}^n$, which satisfies $H(x,1) = G(x)$ and $H(x,0) = F(x)$. The function $G : \mathcal{R}^n \to \mathcal{R}^n$ is a smooth map with known zero points. There are various homotopy functions that are generally used, like the convex homotopy such as

$$H(x, \lambda) = \lambda G(x) + (1-\lambda)F(x),$$

and the Newton homotopy defined by

$$H(x, \lambda) = F(x) - \lambda F(x_1).$$

Once the homotopy function is defined, path following (continuation) methods are applied to track all paths starting at $\lambda = 1$ i.e. at the known solutions of $H(x,1) = 0$ and ending at $\lambda = 0$, converging to the solution of the initial equation $H(x,0) \equiv F(x) = 0$. The path obtained by the sequential solving system of the nonlinear equation

$H(x, \lambda) = 0$ for $\lambda \in [0,1]$ is called the solution path and it is formally defined by the set $H^{-1}(0) = \{(x, \lambda) \mid H(x, \lambda) = 0\}$. Figure 1. represents an example for such a path.

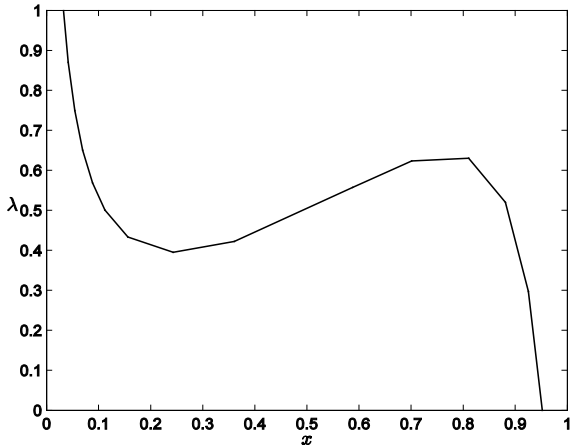

Figure 1. Solution path $H^{-1}(0)$

The solution path can be obtained by successively decreasing the parameter $\lambda$ by a fixed, small increment and solving $H(x, \lambda) = 0$. The drawback of this idea is that it will fail, if turning points of the curve with respect to $\lambda$ are encountered. The remedy to this problem is to parameterize the curve with respect to the arclength $s$. In this way the solution path is defined by the parameterized curve $c(s) = (x(s), \lambda(s)), s \in [0,1]$. Concerning homotopy methods, several questions arise:

1. When is it assured that a curve $c(s) \in H^{-1}(0)$ with $(x_1, 1)$ lying in the curve does exist and is smooth?
2. If such a curve exists, when is it assured that it will intersect the target homotopy level in a finite length?
3. How can such a curve be traced?

The first question is answered by the implicit function theorem, namely, if the Jacobian $\nabla H(x,1)$ has full rank $n$, then a curve $c(s) \in H^{-1}(0)$ with the initial value $c(0) = (x_1, 0)$ and tangent $c'(0) \neq 0$ will exist at least locally i.e. on some open interval around zero [1]. This condition is called regularity condition.

The answer to the second question depends on a particular problem. Generally, it is sufficient to require some boundary condition which will essentially prevent the curve from running to infinity before intersecting the homotopy level $\lambda = 0$.

Solving the system of nonlinear equation $H(x, \lambda) = 0$ is not trivial, because it is underdetermined. Let us recall, it is defined by homotopy function $H : \mathcal{R}^n \times \mathcal{R} \to \mathcal{R}^n$. There are several continuation methods for dealing with underdetermined systems of equation and for tracing the solution path. In order to obtain the solution of the initial system of nonlinear equation, there is no need to determine all points on the zero path. The zero path can be numerically traced by numerical continuation methods. One such path-following method is the predictor-corrector method. The idea is to numerically trace the curve $c(s)$ by generating a sequence of points $u_i = (x_i, \lambda_i), i = 1, 2, \ldots$ along the curve satisfying criterion $\|H(u_i)\| \leq \varepsilon, \varepsilon > 0$. Iteration in the predictor-corrector step consists of two steps, a predictor step, and then a corrector step.

The idea of the predictor step is as follows: the solution of the equation $H(u) = 0$ can be obtained by solving the following initial value problem:

$$H'(u)u = 0, \|u'\| = 1, u(0) = (x_1, 1). \quad (2)$$

Problem (2) can be numerically solved by the Euler method. The predictor step begins at a point near the route and moves in an approximately tangent direction $t$ of the route induced by the Jacobian matrix $H'(u)$. The Euler predictor is defined by

$$v = u + h \cdot t(H'(u)), \quad (3)$$

where $h$ represents a steplength. There are several items of literature like [1] and [3] which propose algorithms for the determination of the steplength.

The purpose of the corrector step is to move from the predicted point to a nearby point on the zero path. This step finds the closest point to the predictor point which is lying on the zero path. It can be done by taking the predictor point as an initial point and performing Newton's method for solving the system of the nonlinear equations $x_{i+1} = x_i - f'(x_i)^{-1} f(x_i)$. As the nonlinear system of equation defined by homotopy function is underdetermined i.e. the matrix of the system is not quadratic, inverse of this matrix does not exist in the classical sense. The Moore-Penrose inverse matrix defined by

$$A^+ = A^T (AA^T)^{-1}$$

is used in Newton's corrector step. The Newton's corrector step is given by

$$w = v - H'(v)^+ \cdot H(v). \quad (4)$$

A version of the Euler – Newton predictor – corrector algorithm can be summarized in following way:

---

**Choose** the initial point $u \in \mathcal{R}^{n+1}$ and algorithm parameters $\varepsilon_1 > 0$ (tolerance for $\lambda$), $\varepsilon_2 > 0$ (the tolerance for solving equation $H(v) = 0$) and the initial steplength $h \in (0,1)$.

**Repeat** until $|\lambda| > \varepsilon_1$ (Euler predictor-step)
$v = u + h \cdot t(H'(u))$.

**Repeat** until $\|H(v)\| > \varepsilon_2$ (Newton corrector-step)
$w = v - H'(v)^+ H(v)$, $v = w$.

**End** corrector loop.

**Choose** new steplength $h \in (0,1)$ according to the asymptotic estimates algorithm given in [1].

**End** predictor loop.

---

*B. Solving NCP by the homotopy method*

A nonlinear complementarity problem with respect to a mapping $f : \mathcal{R}^n \to \mathcal{R}^n$, denoted by NCP($f$), is to find a vector $x \in \mathcal{R}^n$ such that

$$x \geq 0,\ f(x) \geq 0,\ x^T f(x) = 0, \quad (5)$$

where $f(x)$ is a smooth mapping.

Several items of literature including [4] and [9] have documented the basic theory, algorithms and application of the NCP. One of the significant reasons of why complementarity problems are so pervasive in economics is because the concept of complementarity is synonymous with the notion of system equilibrium.

Nowadays there is a wide variety of computational methods for solving complementarity problems. These computational methods include the following:

– extensions of Newton's method for nonlinear equations that replace the direction finding routines with complementarity problems;

– a path search method that uses a generalization of a line search technique;

– quadratic programming-based algorithms that derive extensions of the Gauss-Newton methodology;

– differentiable optimization-based descent methods that reformulate the complementarity relationships as a nonlinear equation or program;

– projection and proximal methods that extend the projected gradient methods;

– smoothing techniques that replace the non-smooth equations with differentiable approximations;

– interior-point methods based on removing inequalities by an interior penalty.

In this paper the NCP will be solved by a path following method. The Euler – Newton predictor – corrector algorithm proposed in the previous section will be applied with special homotopy function proposed in [11].

In order to solve the nonlinear complementarity problem (5), several equivalent formulations of NCP($f$) have been extensively used. One of the formulations is of the form:

$$(x, y) \geq 0,\ F(x, y) = \begin{pmatrix} Xy \\ y - f(x) \end{pmatrix} = 0, \quad (6)$$

where $X = diag(x)$ is a diagonal matrix formed by the components of $x$. Problem (6) can be solved by homotopy method considering a homotopy mapping $H : \mathcal{R}^{2n} \times [0, 1] \to \mathcal{R}^{2n}$ given by

$$H_{w^0}(w, \lambda) = \begin{pmatrix} Xy - \lambda X^0 y^0 \\ y - (1-\lambda) f(x) - \lambda y^0 \end{pmatrix}, \quad (7)$$

where $w = (x, y) \in \mathcal{R}^n \times \mathcal{R}^n$, $\lambda \in [0, 1]$ and an arbitrary positive point is given by $w^0 = (x^0, y^0) \in \mathcal{R}^n_{++} \times \mathcal{R}^n_{++}$. Conditions given in [11] for the convergence of the continuation method with homotopy function (7) are as follows:

1. $f$ is three times continuously differentiable;

2. for any $\{x^k\} \subseteq \mathcal{R}^n_+$, as $k \to \infty$, $\|x^k\| \to \infty$ and $f(x^k) > 0$ when $k > K_0$ for some $K_0 > 0$;

3. the set $S_+ = \{(x, y) \geq 0 : y = f(x)\}$ is non-empty;

Homotopy algorithm defined in this way may diverge for a certain positive fixed point $w^0$, but the parameterized Sard's theorem [11] ensures that the homotopy algorithm will converge with probability one. In the case of divergence, one needs to choose another positive point $w^0$ and to restart the algorithm.

As stated before, homotopy method is a globally convergent method, but a well-chosen starting point can improve the performance of the algorithm. An obvious candidate for the starting point is $w^0 = (x^0, y^0) \in \mathcal{R}^n_{++} \times \mathcal{R}^n_{++}$. In numerical experiments, the algorithm will be tested with several different starting points.

## III. COMPUTATION OF THE ECONOMIC EQUILIBRIA BY THE HOMOTOPY ALGORITHM

Consider a static exchange economy with $I$ consumers and $D$ commodities. Every commodity $d$ has its own positive price $p_d, d = 1, 2, \ldots, D$. The price vector is denoted by $p \in \Re^D$. Each consumer has their own preference defined by the utility function $u_i : X_i \to \Re$, $i = 1, 2, \ldots, I$, where $X_i \subseteq \Re^D$ is the choice set of the consumer, and $\Re^D$ is the commodity space. Vector $x_i \in X_i, i = 1, 2, \ldots, I$ denotes the consumption bundle of the $i$-th consumer. Each consumer is endowed with a vector $w_i \in \mathbb{R}^D, i = 1, 2, \ldots, I$ that is strictly positive. This vector represents the wealth level of the consumer.

An equilibrium of this economy is a price vector $p^* \in \Re^D_{++}$ and an allocation of commodities $(\hat{x}_1, \hat{x}_2, \ldots, \hat{x}_I)$, where $\hat{x}_i \in \Re^D_+$, such that:

1. $\hat{x}_i, i = 1, 2, \ldots, I$ maximizes the preference of the consumer

$$\begin{aligned} & \max u_i(x), \\ & s.t.\ p^{*T} x \leq p^{*T} w_i \\ & x \geq 0 \end{aligned} \quad (8)$$

2. total demand cannot exceed the total endowment $\sum_{i=1}^{I} \hat{x}_i \leq \sum_{i=1}^{I} w_i$.

The solution of the (8) depends from the price vector $p$. It is called demand function. The demand function is mapping $\hat{x} : \Re^D_{++} \to \Re^D_+$ with the property of homogeneity 0, i.e. for all $\lambda > 0$ it holds that $\hat{x}(\lambda p) = \hat{x}(p)$. It is as-sumed that the demand function is continuously differentiable.

Alternatively, the above-described economy can be characterized by the excess demand function $\xi : \mathfrak{R}_{++}^D \to \mathfrak{R}^D$ defined by

$$\xi(p) = \sum_{i=1}^{I} \hat{x}_i - \sum_{i=1}^{I} w_i. \qquad (9)$$

It is assumed that the excess demand function is continuous and homogeneous with degree zero. As a consequence of the homogeneity of degree zero of the demand function in the prices, if $p^*$ represents equilibrium prices, so does $\lambda p^*$ for any scalar $\lambda > 0$. Arguing in this way, one can, without loss of generality, define prices on the positive simplex

$$\triangle_+ = \left\{ p \in \mathfrak{R}_+^D \, \middle| \, \sum_{i=1}^{D} p_i = 1 \right\}.$$

The direct consequence of the homogeneity of degree zero of the excess demand function is that the system of nonlinear equations $\xi(p) = 0$ is underdetermined. By replacing the last equation in the system by the equation $\sum_{i=1}^{D} p_i - 1 = 0$, one obtains system of nonlinear equations which is determined.

In characterization of the equilibrium by the excess demand function, the price $p^* \in \mathfrak{R}_+^D$ is equilibrium if and only if it holds

$$\xi(p^*) \le 0, \quad p^* \ge 0, \quad p^{*T}\xi(p^*) = 0. \qquad (10)$$

The last equation in (10) is known as Walras' law. Problem (10) can be considered as NCP (5) and the homotopy algorithm can be directly applied.

Consider a static production economy by extending exchange economy with production technology. It is assumed that production technology has a constant return to scale. The production set $Y$ exhibits constant return to scale if $y \in Y$ implies that $\alpha y \in Y$ for any scalar $\alpha \ge 0$. Geometrically, production set $Y$ is a cone. The vector $y = (y_1, y_2, \ldots, y_J)$ denotes elementary activities and the production technology is described by an activity analysis matrix $A = [a_{i,j}], i = 1, 2, \ldots, D$, $j = 1, 2, \ldots, J$, where $a_{i,j} \ge 0, (a_{i,j} \le 0)$ denotes output (input). The production defined in this way is called the linear activity model. A price vector $p^* \in \mathfrak{R}_+^D$ and a vector of activity levels $y^* \in \mathfrak{R}_+^J$ constitute a general equilibrium if:

1. $\xi(p^*) - Ay^* \le 0$;

2. $A^T p \le 0$;

3. $p^{*T}(\xi(p^*) - Ay^*) = 0$;

4. $(A^T p^*)^T y = 0$ and

5. $p^* \ge 0, y^* \ge 0$.

Finding general equilibrium which is defined in terms of the above-mentioned definition involves solving the NCP problem. Denoting variables $p$ and $y$ by $x = (p, y)^T$ and $f(x) = (\xi(p) - Ay, A^T p)^T$, the problem of determining the equilibrium can be considered as NCP (5) and it can be solved by the homotopy algorithm. In this setting, the underdetermined system is remedied by replacing the last equation in $A^T p = 0$ with $\sum_{i=1}^{D} p_i - 1 = 0$.

## IV. NUMERICAL RESULTS

In this section the performance of the homotopy algorithm will be studied on well-known test-problems. Test-problems are examples of pure exchange economy and economy with linear production technology. The performance of the homotopy algorithm will be tested with different starting points. For all examples the following is used: $w^0 = 1^T$. During the testing of the performance, the number of the required iteration for meeting stopping rule will be recorded.

The stopping criteria of the interior-point algorithm are given by $\varepsilon_1 = 10^{-6}$ and $\varepsilon_2 = 10^{-5}$ the maximum number of the iteration is given by $Max\_It = 1000$. The initial value for steplength is $h = 0.3$.

**Example 1.** (static exchange economy [12]) Consider a static exchange economy with two consumers and two items of goods. The demand function of the $i$-th consumer is given by

$$\hat{x}_{ij}(p) = \frac{a_{ij}^{1/5} \sum_{k=1}^{2} p_k w_{ik}}{p_j^{1/5} \sum_{k=1}^{2} a_{ik}^{1/5} p_k^{4/5}},$$

where $a_{ij} = 1024$ if $i = j$ and $a_{ij} = 1$ if $i \ne j$. Initial endowments of the consumers are $w_{ij} = 12$ if $i = j$ and $w_{ij} = 1$ if $i \ne j$. The economy defined in this way is example of the economy with several equilibria. Equilibrium points are $p_1^* = (0.5, 0.5)$, $p_2^* = (0.1129, 0.8871)$ and $p_3^* = (0.8871, 0.1129)$. The homotopy algorithm started with several initial points converged to these equilibria. The statistic of the performance of the algorithm is sum-marized in Table I.

TABLE I.
PERFORMANCE OF THE ALGORITHM IN EXAMPLE 1.

|    | Starting point | Iteration No. | Equilibrium |
|----|----------------|---------------|-------------|
| 1. | $(0.01, 0.99)$ | 9             | $p_2^*$     |
| 2. | $(0.1, 0.9)$   | 7             | $p_2^*$     |
| 3. | $(0.4, 0.6)$   | 14            | $p_3^*$     |
| 4. | $(0.5, 0.5)$   | 5             | $p_1^*$     |
| 5. | $(1, 1)$       | 11            | $p_3^*$     |

**Example 2.** (static exchange economy [6]) Consider a static exchange economy with five consumers and ten goods. The $i$-th consumer has an initial endowment $w_i$. The consumer has a demand function of the form

$$\hat{x}_{ij}(p) = \frac{\alpha_i p^T w_i}{p_j^{b_i} \sum_{k=1}^{10} \alpha_{ik} p_k^{1-b_i}}, i=1,\ldots,5, j=1,\ldots,10,$$

where $\alpha_{ik}$ is the demand share parameter for er $i$ and good $k$ and $b_i$ is the elasticity of substitution for consumer $i$. These parameters are given below.

$$\alpha = \begin{bmatrix} 1 & 1 & 3 & 0.1 & 0.1 & 1.2 & 2 & 1 & 1 & 0.7 \\ 1 & 1 & 1 & 1 & 1 & 1 & 1 & 1 & 1 & 1 \\ 9.9 & 0.1 & 5 & 0.2 & 6 & 0.2 & 8 & 1 & 1 & 0.2 \\ 1 & 2 & 3 & 4 & 5 & 6 & 7 & 8 & 9 & 10 \\ 1 & 13 & 11 & 9 & 4 & 0.9 & 8 & 1 & 2 & 10 \end{bmatrix}$$

$b = (2, 1.3, 3, 0.2, 0.6)^T$, and endowments are given by

$$\begin{bmatrix} 0.6 & 0.2 & 0.2 & 20 & 0.1 & 2 & 9 & 5 & 5 & 15 \\ 0.2 & 11 & 12 & 13 & 14 & 15 & 16 & 5 & 5 & 9 \\ 0.4 & 9 & 8 & 7 & 6 & 5 & 4 & 5 & 7 & 12 \\ 1 & 5 & 5 & 5 & 5 & 5 & 5 & 8 & 3 & 17 \\ 8 & 1 & 22 & 10 & 0.3 & 0.9 & 5.1 & 0.1 & 6.2 & 11 \end{bmatrix}$$

This example is the problem of economic equilibrium of higher dimension. The equilibrium price is

$$p^* = (0.187, 0.109, 0.099, 0.043, 0.117,$$
$$0.077, 0.117, 0.102, 0.099, 0.049).$$

The homotopy algorithm converged to equilibrium. The performance of the algorithm for different starting points is given in Table II.

TABLE II.
PERFORMANCE OF THE ALGORITHM IN EXAMPLE 2.

|   | Starting point | Iteration No. |
|---|---|---|
| 1. | $1^T$ | 21 |
| 2. | $\frac{1}{10} \cdot 1^T$ | 15 |

**Example 3.** (production economy [12]) Consider a production economy with two consumers and four items of goods. The demand function is given by the correspondence

$$\hat{x}_{ij}(p) = \frac{a_{ij} \sum_{k=1}^{4} p_k w_{ik}}{p_j \sum_{k=1}^{4} a_{ik}}, i=1,2, j=1,\ldots,4$$

where initial endowments and parameter $a$ are given by matrices

$$w = \begin{bmatrix} 0 & 0 & 10 & 0 \\ 0 & 0 & 0 & 20 \end{bmatrix}, a = \begin{bmatrix} 0.8 & 0.2 & 0 & 0 \\ 0. & 0.9 & 0 & 0 \end{bmatrix}.$$

The production technology is specified by an activity matrix $A$

$$\begin{bmatrix} -1 & 0 & 0 & 0 & 3 & 5 & -1 & -1 \\ 0 & -1 & 0 & 0 & -1 & -1 & 5 & 5 \\ 0 & 0 & -1 & 0 & -1 & -1 & -1 & -4 \\ 0 & 0 & 0 & -1 & -1 & -4 & -3 & -1 \end{bmatrix}.$$

The problem of finding the equilibrium is an example of economy with equilibrium which is not unique. Reference [12] pointed out that this problem has three equilibria. The first equilibrium is $p_1^* = (0.25, 0.25, 0.25, 0.25)$ with activity level $y_1^* = (0,0,0,0,5,0,5,0)$. The second equilibrium is $p_2^* = (0.2500, 0.2222, 0.3611, 0.1667)$ with the activity level $y_2^* = (0,0,0,0,5.1806, 0.3611, 4.4583, 0)$ and finally, the third equilibrium point is $p_3^* = (0.2500, 0.2708, 0.1667, 0.1190)$ with optimal allocation $y_3^* = (0,0,0,0,4.3690, 0, 5.1548, 0.1190)$. For different starting points the algorithm converged to a second equilibrium.

TABLE III.
PERFORMANCE OF THE ALGORITHM IN EXAMPLE 3.

|   | Starting point | Iteration No. |
|---|---|---|
| 1. | $p_0 = 1^T, y_0 = 1^T$ | 28 |
| 2. | $p_0 = \frac{1}{4} 1^T, y_0 = 1^T$ | 24 |
| 3. | $p_0 = (0.1, 0.7, 0.001, 0.19)^T, y_0 = 1^T$ | 26 |
| 4. | $p_0 = (0.1, 0.7, 0.01, 0.19)^T$ $y_0 = (0,0,0,0,1,3,0.10.3)^T$ |  |

This point is possibly obtained because it is the local minimum.

**Example 4.** (production economy [8]) Consider a static production economy with one consumer, one producer and three items of goods. The consumer has a demand function of the form

$$\hat{x}_i(p) = \frac{a_i(w_2 \cdot p_2 + w_3 \cdot p_3)}{p_i}, i=1,2,3,$$

and initial endowment $w = (0,5,3)$. The firm has technology matrix $A = (1,-1,-1)$. The parameter of the demand function is given by $a = (0.9, 0.1, 0)$. The algorithm has been tested with several starting points and always converged to the normalized equilibrium allocation $p^* = (0.5000, 0.0833, 0.4167)$ and to the activity level $y^* = 3$. The iteration number needed for reaching the equilibrium was between 14 and 25 depending on the initial point.

**Example 5.** (production economy [6]) Consider a static production economy with five consumers, eight activity sectors and six items of goods. Each consumer has an initial endowment $w_i$. The consumer has a demand function of the form

$$\hat{x}_{ij}(p) = \frac{a_{ij} p^T w_i}{p_j^{b_i} \sum_{k=1}^{6} a_{ik} p_k^{1-b_i}}, i=1,\ldots,5, j=1,\ldots,6,$$

where $a$ is the demand share parameter for consumer $i$, and $b_i$ is the elasticity of substitution for consumer $i$. The values of parameters and endowment are given in [6]. The equilibrium allocation is given by normalized price vector $p^* = (0.2203, 0.2511, 0.1610, 0.0549, 0.1061, 0.2066)$ and $y^* = (0^{1\times 6}, 0.4635, 0, 3.9392, 0.0060, 0, 0, 0.4383, 0)$.

The homotopy algorithm reached this equilibrium with several initial points in $28 - 37$ iteration.

**Example 6.** (production economy [6]) Consider a static production economy with four consumers, twenty-six activity sectors and fourteen items of goods. Each consumer has an initial endowment $w_i$. The consumer has a demand function of the form

$$\hat{x}_{ij}(p) = \frac{a_{ij} p^T w_i}{p_j}, i=1,\ldots,4, j=1,\ldots,14,$$

where $a$ is the demand share parameter for consumer $i$. The values of parameters and endowments are given in [6]. The equilibrium allocation is given by price vector

$p^* = (0.06215, 0.05833, 0.09545, 0.07145, 0.06585,$
$0.06245, 0.06890, 0.09811, 0.09024, 0.07956, 0.05620,$
$0.06201, 0.03652, 0.09279),$

and production allocation

$y^* = (4.7923, 0_{2\times 1}, 51.9714, 4.0414, 0_{3\times 1}, 30.5004,$
$21.1848, 36.8945, 28.0286, 0, 44.0441, 23.6464,$
$0, 25.6427, 0, 12.0530, 0_{3\times 1}, 47.2847, 0_{3\times 1}).$

For different starting points, the algorithm converged to equilibrium allocation. The iteration number needed for reaching the equilibrium was between 445 and 493 iteration.

### III. SUMMARY AND CONCLUSION

There are many methods for computation of equilibria in static exchange and static production equilibrium models. When the functional form of the excess demand function is available, the definition of the equilibrium can be considered as a NCP. The main idea is the computation of economic equilibria by solving the NCP. In the continuation method the new homotopy functions is used.

Homotopy methods are extensively used in computing economic equilibria, because they can be used in constructive proofs of economic equilibrium existence. Literature [10] argues against using homotopy methods for computing economic equilibria, because they have difficulties with solving medium to large scale equilibrium models due to the complexity of the algorithm. Considering numerical examples introduced in the previous section, homotopy algorithm is robust algorithm and does not have problems with medium scale equilibrium problems. Example 6. confirms this statement. Compared to interior-point algorithm proposed in [10] and analyzed in [13], it is more robust.